# An Alternate Graphical Representation of Periodic table of Chemical Elements

Mohd Abubakr[1],
Microsoft India (R&D) Pvt. Ltd,
Hyderabad, India.
mohdabubakr@hotmail.com

## Abstract

Periodic table of chemical elements symbolizes an elegant graphical representation of symmetry at atomic level and provides an overview on arrangement of electrons. It started merely as tabular representation of chemical elements, later got strengthened with quantum mechanical description of atomic structure and recent studies have revealed that periodic table can be formulated using $SO(4,2) \otimes SU(2)$ group. IUPAC, the governing body in Chemistry, doesn't approve any periodic table as a standard periodic table. The only specific recommendation provided by IUPAC is that the periodic table should follow the 1 to 18 group numbering. In this technical paper, we describe a new graphical representation of periodic table, referred as 'Circular form of Periodic table'. The advantages of circular form of periodic table over other representations are discussed along with a brief discussion on history of periodic tables.

## 1. Introduction

The profoundness of inherent symmetry in nature can be seen at different depths of atomic scales. Periodic table symbolizes one such elegant symmetry existing within the atomic structure of chemical elements. This so called 'symmetry' within the atomic structures has been widely studied from different prospects and over the last hundreds years more than 700 different graphical representations of Periodic tables have emerged [1]. Each graphical representation of chemical elements attempted to portray certain symmetries in form of columns, rows, spirals, dimensions etc. Out of all the graphical representations, the rectangular form of periodic table (also referred as Long form of periodic table or Modern periodic table) has gained wide acceptance. However, International Union of Pure and Applied Chemistry (IUPAC), does not approve any form as periodic table as a standard form [2]. An IUPAC approved version of periodic table doesn't exist. With regard to periodic table IUPAC, only recommends the use to 1 to 18 for group numbering as standard nomenclature.

---

[1] The views expressed in this paper are the views of the author and do not necessarily reflect the views of Microsoft India (R&D) Pvt. Ltd.

## 2. Brief History of Periodic Tables

Prior to Mendeleev's pioneering work in creation of Periodic table, several models such as Dobereiner traids(1817), Gmelin(1827), Pettenkofer(1850), Beguyer de Chancourtois' spiral(1862), Newlands' octaces(1865), Olding tables(1865), Meyer(1870) were known [1]. This was the ad-hoc phase of graphical representations of chemical elements. Considering that from 1815 to 1865, the number of known chemical elements almost doubled, discovery of every new element came along with new patterns and theoretical speculations. Speculations emerged due to the fact that each new element not only had certain unique chemical properties but also resembled with other elements in certain areas. In 1865, Mendeleev pioneered the work of classifying all the known elements under a periodic law of atomic weights. According to Mendeleev's table, all the elements are represented under a simple tabular format with increasing order of the weights and arranged the elements of analogous nature in columns. Mendeleev was successful in giving the first complete graphical representation of chemical elements and also succeeded in predicting new elements purely based on symmetric properties in the tables. Prediction of new elements based on symmetry led to the phenomenal success of graphical representation and opened doors in exploring the symmetries within atomic elements.

```
                            Ti=50      Zr=90      ?=180
                            V=51       Nb=94      Ta=182
                            Cr=52      Mo=96      W=186
                            Mn=55      Rh=104,4   Pt=197,4
                            Fe=56      Ru=104,4   Ir=198
                            Ni=Co=59   Pd=106,6   Os=199
        H=1                 Cu=63,4    Ag=108     Hg=200
            Be=9,4  Mg=24   Zn=65,2    Cd=112
            B=11    Al=27,4 ?=68       Ur=116     Au=197?
            C=12    Si=28   ?=70       Sn=118
            N=14    P=31    As=75      Sb=122     Bi=210?
            O=16    S=32    Se=79,4    Te=128?
            F=19    Cl=35,5 Br=80      J=127
        Li=7 Na=23  K=39    Rb=85,4    Cs=133     Tl=204
                    Ca=40   Sr=87,6    Ba=137     Pb=207
                    ?=45    Ce=92
                    ?Er=56  La=94
                    ?Yt=60  Di=95
                    ?In=75,6 Th=118?
```

Fig. 1: Mendeleev's Periodic table, elements are arranged in increasing order of their weight and analogous elements are arranged in rows and columns.

In early 20th century, Quantum mechanical properties of nature revealed that number of electrons in an element determines its chemical properties rather than atomic weights. The number of electrons within an atom was then termed as '*atomic number*', represented by number '*Z*'. According to [1], three reasons that established atomic number as fundamental are

(i)     Rutherford's formula for the differential cross-section in the scattering of alpha particles (1911)

(ii)    Moseley's formula for frequency of X-ray lines (1913) and

(iii)   Bohr-Balmer formula for the energy levels of hydrogen-like atom (1913).

These discoveries along with other advances in quantum mechanics, led to the design of modern periodic table with a separate block for transition metals and rare-earth metals. Further advances in quantum mechanics confirmed that chemical properties of an atom are indeed based on the total number of electrons and the number of valence electrons. Based on the value of Z, four quantum numbers were defined to describe atomic properties of elements. The overall behavior of the atom including chemical properties depend on these four quantum numbers namely, principle, azimuthal, magnetic and spin quantum number of the outermost electron. Quantum numbers are defined for each electron available within an atom. Principle quantum number for an electron determines the shell number it belongs to, whereas azimuthal number provides information about sub-shell. Based on the quantum numbers of outermost electron of outermost shell, the chemical properties of an atom are determined. Magnetic and spin quantum number doesn't directly indulge in chemical properties however plays an important role in electronic configuration of elements. According to Aufbau's principle, electrons should occupy the lower-most electronic sub-shell available. This forms the basis for representation of elements within Modern form of periodic table, also referred as "long form of periodic table".

Energy of the electron increases with (n+l), where n, l are principal and azimuthal quantum number. If two electrons within an atom have equal (n+l) value, then electron with higher value of 'n' has higher energy. This rule is referred as Madelung's rule. Madelung's rule has high significance in determining the group and position of an element in Periodic table. For excellent description about quantum mechanical aspects and history of periodic tables, refer [8].

The long form of Periodic table is based on the electronic configuration of the valence electrons of elements. It consists of 18 vertical columns called as groups and 7 horizontal rows called as periods. Also to emphasize on the importance of the outer shell electrons the table is divided into blocks namely s, p, d and f blocks. With many advantages over Mendeleev's Periodic Table, the Modern form of periodic table has gained wide popularity. Various properties that Modern Periodic table depicts graphically are size, chemical behavior, atomic radius, ionization energy, electron affinity, electro negativity metallic character, binding energies, diagonal relationships etc.

Fig 2: Long form of periodic table

Recently, Maurice Kibler proposed the application of group SO(4,2)⊗SU(2) to the periodic table of chemical elements [5-6]. If true, this is an important discovery as it opens doors for revealing new symmetries within elements.

Fig. 3: SO(4,2)⊗SU(2) group representation of Periodic Table (X – non discovered element)

# 3. Circular form of Periodic table

The major disadvantage of the Modern Periodic table is its shape itself. Rectangular arrangement of elements puts constraints in describing the properties of the elements such as shape, atomic structure, nucleus, etc. With the developments in the nucleon shell model it is far more informative to have a complete periodic table showing both nucleon as well as electronic structure. To incorporate additional properties in a form of arrangement of elements, we describe a new graphical representation of periodic table[2], which is referred as 'Circular form of Periodic Table'. The aim of this circular form of table is make a student understand the atomic structure and its properties alongside the periodic table. While this new graphical form does include all the properties of the Modern Periodic table, it also includes additional information from atomic structure point of view.

## 3.1 *Arrangement of groups and periods*

Circular form of periodic tables includes 18 sectors and 7 shells. Each sector represents a group, whereas each shell represents the period. Hydrogen and Helium are given the special position in the first shell. IUPAC recommendations are followed for group numbering. The 7 shells present in the graphical representation symbolizes electron orbits around the nucleus. Similar to long form of periodic table, the lanthanides and actinides are isolated from the main table and placed below the circular table as an arc. The s, p, d and f blocks of long form of periodic table are retained, albeit the blocks contain group of sectors than columns.

## 3. 2 *Position of Hydrogen and Helium*

The position of Hydrogen and Helium has been debatable in long form of periodic table. Based on the definition of groups, Hydrogen can be placed as alkali metal or as a halogen. From the outer-most electronic configuration point of view, Hydrogen has one 1s electron, therefore it should be placed in 1st group. However, it needs only one electron to attain inert configuration, it can also be placed in 17th (halogen) group. Given that, chemically hydrogen behaves both as a halogen as well as alkali metal, the position of Hydrogen is uncertain. Helium is another element whose position is unsatisfactory. The outermost configuration of Helium is $1s^2$, therefore it should be placed in 2nd group however the chemical properties of Helium are equivalent to that of inert gases, positioned at 18th group. In the long form of periodic table, precedence is given to chemical properties of Helium by placing it in 18th group. Note that, all the other elements in the 18th group have $2p^6$ as their outermost electronic configuration.

---

[2] This graphical representation was first designed by the author in 2001. It was later posted on Wikipedia by the author in 2006. (http://commons.wikimedia.org/wiki/File:Circular_form_of_periodic_table.svg)

Circular form of periodic table captures an interesting fact about position of Hydrogen and Helium. Consider for example, the position of Hydrogen needs to satisfy two conditions.
1. Behavior of Hydrogen as an 1st group element (alkali)
2. Behavior of Hydrogen as a 17th group element (halogen)

In terms of position, we reword the above conditions as
1. Hydrogen should be placed in the group prior to 2nd group to satisfy alkali properties.
2. Hydrogen should be placed in the group prior to 18th group to satisfy halogen properties.

We obtain the position of the Hydrogen in periodic table as follows,
1. A line L1 is drawn across the table such that, it passes through the center of the nucleus and intersects 2nd group sector.
2. Another line L2 is drawn across the table such that, it passes through the center of the nucleus and intersects 18th group sector.
3. Lines L1 and L2, when extended upper side of the nucleus, create a diverging sector. Two arcs are drawn on this sector, such that they represent the first shell across the nucleus.
4. The area enclosed in the first shell on the sector is the position for Hydrogen.

The position of Helium in the periodic table is obtained similarly by extending lines from 1st group and 17th group. However, the size of the first shell drawn for Helium is slightly larger compared to Hydrogen.

## *3.3 Advantages of Circular form of periodic table*

1. Graphical representation of the arrangement of chemical elements provides a visual aid in recognizing the symmetry and patterns. A Circular model with nucleus at the center depicts the shape of the atom obtained from Bohr's model [3, 7], hence incorporating the "shape" representation which was absent in Long form of periodic table.
2. Atomic radius increases across with inner shell to outer shell. The lesser the atomic radius, closer is the outer most electron to the nucleus. This property is captured in the circular form of periodic table.
3. Properties such as ionization energy, electron affinity, electro negativity metallic character, binding energies, diagonal relationships holds good for Circular form of Periodic Table.
4. Special position is assigned for Hydrogen and Helium considering their atomic properties. Also, both Hydrogen and Helium are placed closed to nucleus unlike other elements in the graphical representation. This is in accord with properties of hydrogen and helium.

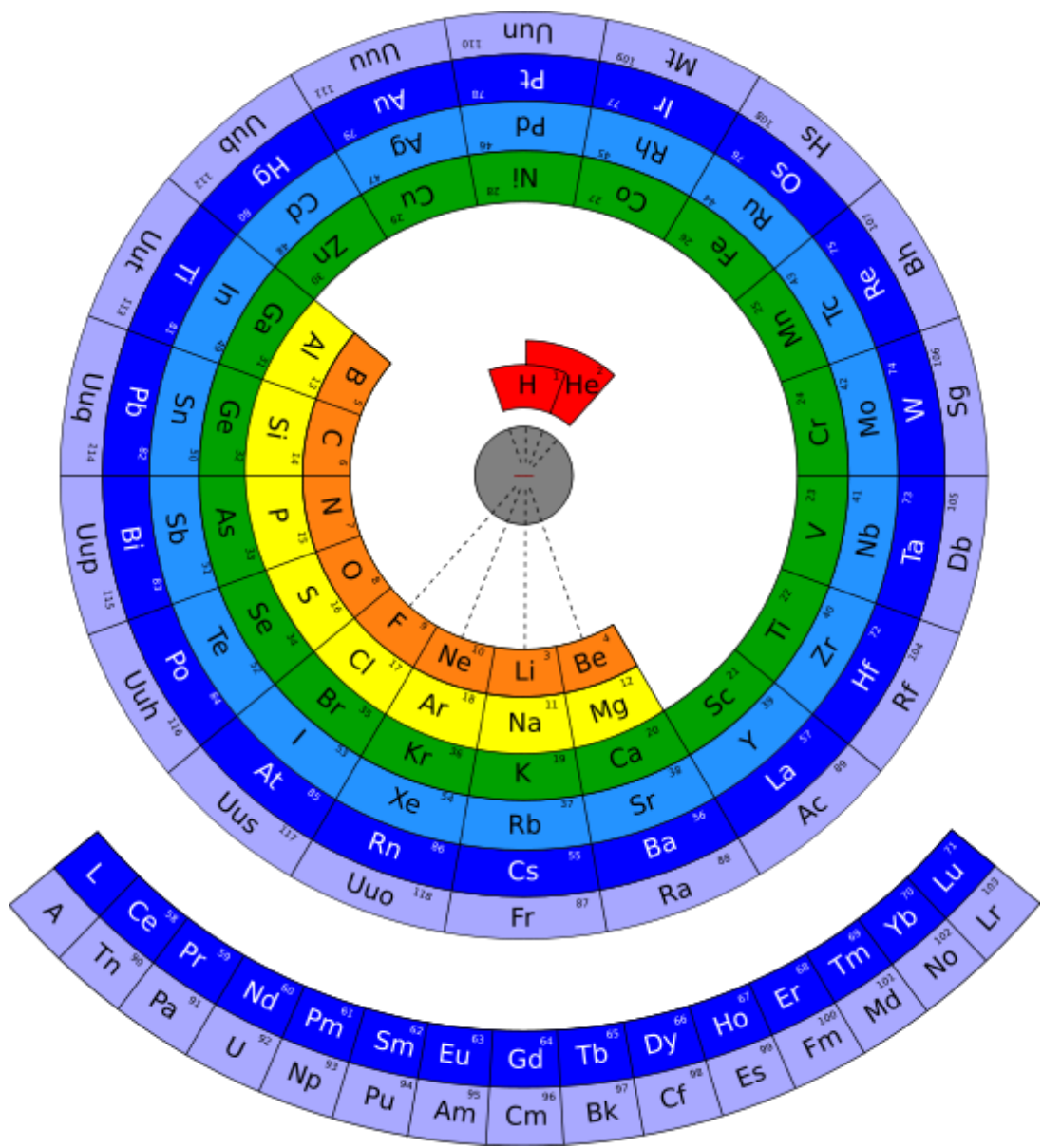

Fig. 4: Circular form of Periodic table of Elements

## 4. Conclusion

A new graphical representation of periodic table referred as "circular form of periodic table" is described in this paper. This new graphical representation abides the recommendation given by IUPAC about group numbering. Given that essence of periodic table is to capture as many properties of elements as possible while retaining the simple design, circular form of periodic table offers several advantages over other widely known forms of periodic tables. We humbly request the faculty members at various educational institutes to use this new form of periodic table in teaching. It will help the students in visualizing the atomic shape of the elements along with the electronic configuration over other periodic table designs.